\documentclass[twocolumn, floatfix, showpacs, eqsecnum, superscriptaddress, prb, aps, 10pt]{revtex4}
\usepackage{graphicx}
\usepackage{amsmath}
\usepackage{amssymb}
\usepackage{bm}
\usepackage{array}
\usepackage{color}

\DeclareMathOperator{\tr}{Tr}

\begin{document}

\title{Adaptive tuning of Majorana fermions in a quantum dot chain.}
\author{Ion C. Fulga}
\affiliation{Instituut-Lorentz, Universiteit Leiden, P.O. Box 9506, 2300 RA Leiden, The Netherlands}
\author{Arbel Haim}
\affiliation{Department of Condensed Matter Physics, Weizmann Institute of Science, Rehovot, 76100, Israel}
\author{Anton R. Akhmerov}
\affiliation{Instituut-Lorentz, Universiteit Leiden, P.O. Box 9506, 2300 RA Leiden, The Netherlands}
\affiliation{Department of Physics, Harvard University, Cambridge, MA 02138}
\author{Yuval Oreg}
\affiliation{Department of Condensed Matter Physics, Weizmann Institute of Science, Rehovot, 76100, Israel}

\date{\today}
\begin{abstract}
We suggest a way to overcome the obstacles that disorder and high density of states pose to the creation of unpaired
Majorana fermions in one-dimensional systems.
This is achieved by splitting the system into a chain of quantum dots, which are then tuned to the conditions under which the chain
can be viewed as an effective Kitaev model, so that it is in a robust
topological phase with well-localized Majorana states in the outermost dots. The tuning
algorithm that we develop involves controlling the gate voltages and
the superconducting phases. Resonant Andreev spectroscopy allows us to make the tuning adaptive, so that each pair of dots may be
tuned independently of the other. The calculated quantized zero bias conductance serves then as a natural proof of the topological nature of the tuned phase.
\end{abstract}
\pacs{74.45.+c, 74.78.Na, 73.63.Kv, 03.65.Vf}
\maketitle

\section{Introduction}
\label{intro}

Majorana fermions are the simplest quasiparticles predicted to have
non-Abelian statistics.\cite{Ali12, Bee11} These topologically protected states
can be realized in condensed matter systems, by making use of a combination of
strong spin-orbit coupling, superconductivity, and broken time-reversal
symmetry.\cite{Sau10,Ore10,Lut10} Recently, a series of experiments have reported the
possible observation of Majorana fermions in semiconducting nanowires,\cite{Mou12,
Den12, Das12, Rok12} attracting much attention in the condensed matter
community.

Associating the observed experimental signatures exclusively with these non-Abelian 
quasiparticles, however, is not trivial. The most straightforward signature, the
zero bias peak in Andreev conductance\cite{Bol07, Law09} is not unique to
Majorana fermions, but can appear as a result of various physical mechanisms,\cite{Sas00, Pik12,
Fle10, Kel12, Tew12, Pie12, Liu12} such as the Kondo effect or weak
anti-localization. It has also been pointed out that disorder
has a detrimental effect on the robustness of the topological phase, since in
the absence of time-reversal symmetry it may close the induced
superconducting gap.\cite{And59} This requires experiments performed with very clean
systems. Additionally, the presence of multiple transmitting modes reduces the 
amount of control one has over such systems,\cite{Pot10, Sta11, Bro11, Rie12}
and the contribution of extra modes to conductance hinders the observation of Majorana fermions.\cite{Wim11} Thus, nanowire experiments
need setups in which only few modes contribute to conductance.

\begin{figure}[h!]
 \includegraphics[width=0.9\linewidth]{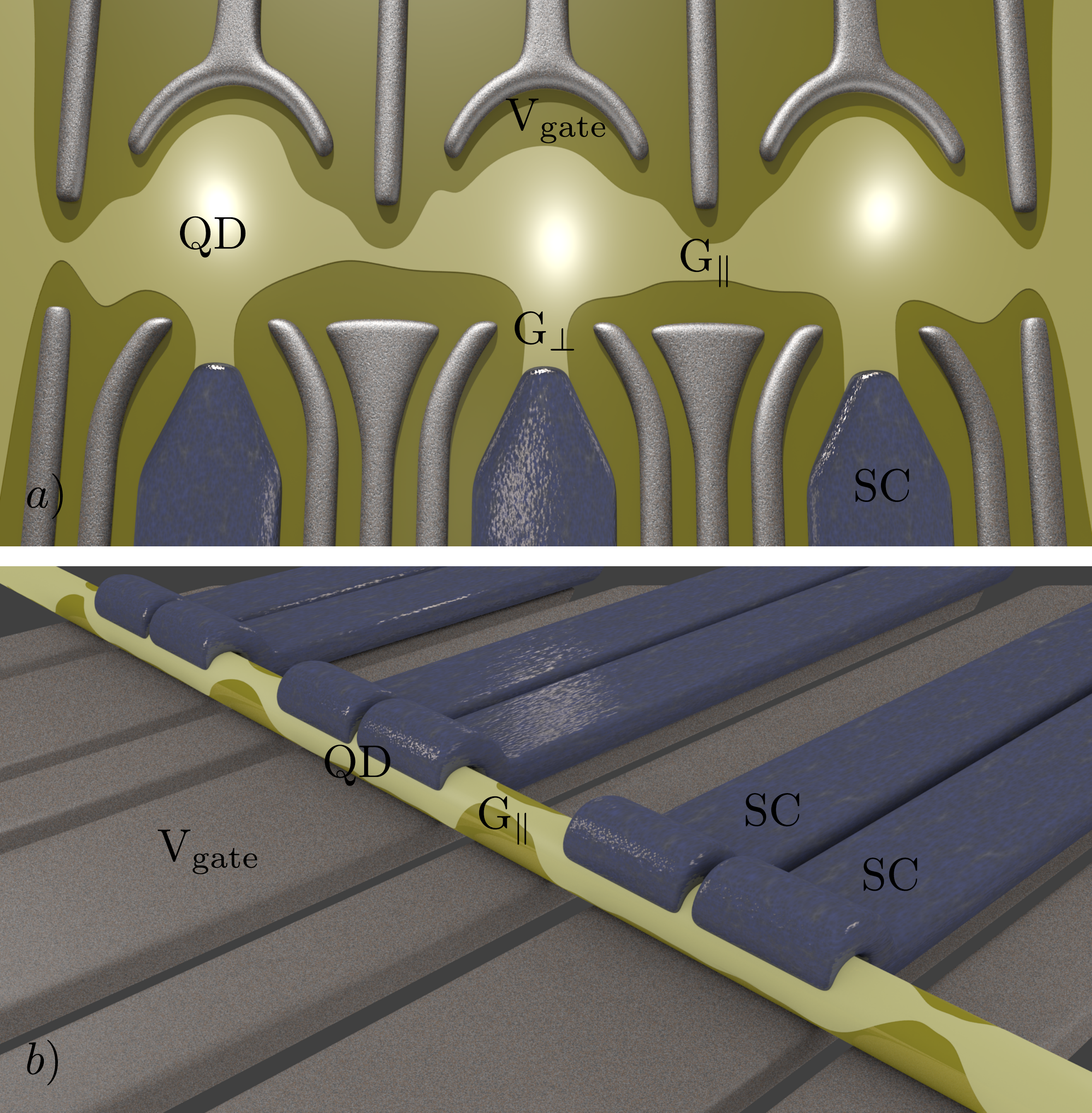}
 \caption{Examples of systems allowing implementation of a Kitaev chain.
Panel (a): a chain of quantum dots in a 2DEG. The QDs are connected to each
other, and to superconductors (labeled SC), by means of quantum point
contacts. 
The first and the last dots are also coupled to external leads.
The normal state  conductance of quantum point contacts (QPCs) between adjacent dots or between the end
dots and the leads is $G_\parallel$, and of the QPCs linking a dot to a
superconductor is $G_\perp$. The confinement energy inside each QD can be
controlled by varying the potential $V_{\rm gate}$. Panel (b): Realization
of the same setup using a nanowire, with the difference that each dot is
coupled
to two superconductors in order to control the strength of the superconducting
proximity effect without the use of QPCs.\label{fig:system}}
\end{figure}

In this work we approach the problem of realizing systems in a non-trivial
topological phase from a different angle. We wish to emulate the Kitaev chain
model\cite{Kit01} which is the simplest model exhibiting unpaired Majorana
bound states. The proposed system consists of a chain of
quantum dots (QDs) defined in a two-dimensional electron gas (2DEG) with spin orbit coupling, 
in proximity to superconductors and subjected to an external magnetic
field. Our geometry enables us to control the parameters of the system to a
great extent by varying gate potentials and superconducting phases.  We will
show how to fine tune the system to the so-called ``sweet spot'' in parameter
space, where the Majorana fermions are well-localized at the ends of the
system, making the topological phase maximally robust. A sketch of our
proposed setup is presented in Fig.~\ref{fig:system}a).

The setup we propose and the tuning algorithm is not restricted solely to
systems created in a two-dimensional electron gas. The essential components are
the ability to form a chain of quantum dots and tune each dot separately. In
semiconducting nanowires the dots can be formed from wire segments separated by
gate-controlled tunnel barriers,
and all the tuning can be done by gates, except for the coupling to a
superconductor. This coupling, in turn, can be controlled by coupling two
superconductors to each dot and applying a phase difference to these superconductors. The layout of a nanowire implementation 
of our proposal is shown in Fig.~\ref{fig:system}b).

This geometry has the advantage of eliminating many of the problems mentioned
above. By using single level quantum dots, and also quantum point contacts (QPC)
in the tunneling regime, we solve issues related to multiple transmitting
modes. Additional problems, such as accidental closings of the induced
superconducting gap due to disorder, are solved because our setup allows us to tune the system to a point where the topological phase
is most robust, as we will show.

We present a step-by-step tuning procedure which follows the behavior of the
system in parallel to that expected for the Kitaev chain. As feedback required to control every step we
use the resonant Andreev conductance, which allows to track the evolution of the system's energy
levels. We expect that the step-by-step structure of the tuning algorithm should eliminate the
large number of non-Majorana explanations of the zero bias peaks.

A related layout together with the idea of simulating a Kitaev chain was proposed
recently by J. D. Sau and S. Das Sarma.\cite{Sau12} Although similar in nature,
the geometry which we consider has several advantages. First of all, coupling the superconductors to the
quantum dots in parallel, allows us to not rely
on crossed Andreev reflection. More importantly, being able to control inter-dot coupling
separately from all the other properties allows to address each dot or each segment of the chain electrically.
This can be achieved by opening all the QPCs except for the ones that contact the desired dots.

This setup can also be extended to more complicated geometries
which include T-junctions of such chains. Benefiting from the high tunability of
the system and the localization of the Majorana fermions, it might then be
possible to implement braiding\cite{Ali11, Sau11} and demonstrate their non-Abelian nature.

The rest of this work is organized as follows. In section \ref{kitaev} we
briefly review a generalized model of Kitaev chain, and identify the "sweet
spot" in parameter space in which the Majorana fermions are the most
localized. The system of coupled quantum dots is described in section
\ref{system}. For the purpose of making apparent the resemblance of the system
to the Kitaev chain, we present a simple model which treats each dot as having
a single spinful level. We then come up with a detailed tuning procedure
describing how one can control the parameters of the simple model, in order to
bring it to the desired point in parameter space. In section \ref{perform_tune}
our tuning prescription is applied to the suggested system of a chain of QDs
defined in a 2DEG, and it is shown using numerical simulations that at the end
of the process the system is indeed in a robust topological phase. We conclude
in section \ref{conc}.

\section{Generalized Kitaev chain}
\label{kitaev}

In order to realize unpaired Majorana bound states, we start from the Kitaev
chain\cite{Kit01} generalized to the case where the on-site
energies as well as the hopping terms are not uniform and can vary from site to
site. The generalized Kitaev chain Hamiltonian is defined as
\begin{align}
    H_\text{K} = \sum_{n=1}^{L-1}  \Big[ \Big( & t_n e^{i \theta_n} a^\dag_{n+1}a_n +
                 \Delta_n e^{i \phi_n} a^\dag_{n+1}a^\dag_n  \nonumber\\
     &+ {h.c.} \Big) +\varepsilon_n a^\dag_n a_n \Big] ,\label{eq:HKitaev}
\end{align}
where $a_n$ are fermion annihilation operators, $\varepsilon_n$ are the on-site
energies of these fermions, $t_n \exp(i \theta_n)$ are the hopping
terms, and $\Delta_n\exp(i\phi_n)$ are the p-wave pairing terms.

The chain supports two Majorana bound states localized entirely on the first
and the last sites, when (i): $\varepsilon_n=0$, (ii): $\Delta_n = t_n$, and (iii) $\phi_{n+1} - \phi_n - \theta_{n+1} - \theta_n = 0$.
The larger values of $t_n$ lead to a larger excitation gap.
The condition (iii) is equivalent, up to a gauge transformation, to the case
where the hopping terms are all real, and the phases of the p-wave terms are
uniform. The energy gap separating the Majorana modes from the first excited
state then equals
\begin{equation}
E_\text{gap}=2\min\left\{t_n\right\}_n.\label{eq:Hk_Eg}
\end{equation}

The above conditions (i)--(iii), constitute the ``sweet spot'' in
parameter space to which we would like to tune our system. Since all of these conditions are local
and only involve one or two sites, our tuning procedure
includes isolating different parts of the system
and monitoring their energy levels. For that future purpose we will use the expression for excitation energies of a
chain of only two sites with $\varepsilon_1=\varepsilon_2=0$:
\begin{equation}
E_{12} = \pm (t_1 \pm \Delta_1).\label{eq:twosites}
\end{equation}

\section{System description and the tuning algorithm}
\label{system}

The most straightforward way to emulate the Kitaev chain is to create an array of spinful quantum
dots, and apply a sufficiently strong Zeeman field such that only one spin state stays close to the Fermi
level. Then the operators of these spin states span the basis of the Hilbert space of the Kitaev chain.
If we require normal hopping between the dots and do not utilize crossed Andreev reflection,
then in order to have both $t_n$ and $\Delta_n$ nonzero we need to break the particle number conservation and
spin conservation. The former is achieved by coupling each dot to a superconductor, the latter can be achieved
by spatially varying Zeeman coupling,\cite{Cho11, Kja12} or more conventionally by using a material with
a sufficiently strong spin-orbit coupling. Examples of implementation of such a
chain of quantum dots in a two dimensional electron gas and in semiconducting
nanowires are shown in Fig.~\ref{fig:system}.

We neglect all the levels in the dots except for the one closest to the Fermi 
level, which is justified if the level spacing in the dot is larger than all
the other Hamiltonian terms. We neglect the Coulomb blockade, since we assume
that the dot is strongly coupled to the superconducting lead. The general form
of the BdG Hamiltonian describing such a chain of spinful single-level dots is
then given by:
\begin{equation}\label{eq:H_S}
\begin{split}
H_\text{S} &= \sum_{n,s,s'} \left(\mu_n \sigma_0 +V_z \sigma_z \right) c_{n,s}^\dag c_{n,s'} \\
&+ \frac{1}{2} \left(\Delta_{\text{ind},n} e^{i\Phi_n} i\sigma_y c_{n,s}^\dag c_{n,s'}^\dag + \textrm{h.c.}\right)\\
&+ \left(w_n e^{i \bm{\lambda}_n \bm{\sigma}} c_{n,s}^\dag c_{n+1,s'} + \textrm{h.c.} \right),
\end{split}
\end{equation}
where $c_{n,s}^{\dagger}$ and $c_{n,s}$ are creation and annihilation operators
of a fermion with spin $s$ in the $n$-th dot, and $\sigma_i$ are Pauli
matrices in spin space. The physical quantities entering this Hamiltonian are
the chemical potential $\mu_n$, the Zeeman energy $V_z$, the proximity-induced
pairing $\Delta_{\textrm{ind},n} \exp(i \Phi_n)$, and the inter-dot hopping
$w_n$. The vector $\bm{\lambda}_n$ characterizes the amount of spin rotation
happening during a hopping between the two neighboring dots (the spin rotates
by a $2|\lambda|$ angle). This term may be
generated either by a spin-orbit coupling, or by a position-dependent spin
rotation, required to make the Zeeman field point in the local
$z$-direction.\cite{Bra10, Cho11, Kja12} The induced pairing in each dot
$\Delta_{\text{ind},n}\exp(i\Phi_n)$ is not to be confused with the p-wave
pairing term $\Delta_n\exp(i\phi_n)$ appearing in the Kitaev chain Hamiltonian
\eqref{eq:HKitaev}.

In order for the dot chain to mimic the behavior of the Kitaev chain in the 
sweet spot, each dot should have a single fermion level with zero energy, so
that $\varepsilon_n = 0$. Diagonalizing a single dot Hamiltonian yields the
condition for this to happen:
\begin{equation}
\mu_n=\sqrt{V_z^2-\Delta_{\text{ind},n}^2}.\label{eq:Mu}
\end{equation}
When this condition is fulfilled, each dot has two fermionic excitations
\small
\begin{gather}
a_n = \frac{e^{i\frac{\Phi_n}{2}}}{\sqrt{2V_z}}\left(\sqrt{V_z-\mu_n}\,\, c_{n\uparrow}^\dag-e^{-i\Phi_n}\sqrt{V_z+\mu_n}\,\,c_{n\downarrow}\right)\\
\,\,\,b_n = \frac{e^{i\frac{\Phi_n}{2}}}{\sqrt{2V_z}}\left(\sqrt{V_z-\mu_n}\,\, c_{n\downarrow}^\dag+e^{-i\Phi_n}\sqrt{V_z+\mu_n}\,\,c_{n\uparrow}\right).
\label{eq:transformation}
\end{gather}
\normalsize
The energy of $a_n$ is zero, the energy of $b_n$ is $2 V_z$. If the hopping 
is much smaller than the energy of the excited state, $w_n \ll V_z$, we may
project the Hamiltonian \eqref{eq:H_S} onto the Hilbert space spanned by $a_n$.
The resulting projected Hamiltonian is
identical to the Kitaev chain
Hamiltonian of Eq.~\eqref{eq:HKitaev}, with the following effective parameters:
\begin{subequations}
\begin{align}
\varepsilon_n &= 0,\\
t_ne^{i\theta_n} &= w_n\left(\cos\lambda_n + i\sin\lambda_n \cos\rho_n\right)\times \notag\\
\bigg[&\sin\left(\alpha_{n+1}+\alpha_n\right)\cos(\delta\Phi_n/2) \\
+ i&\cos\left(\alpha_{n+1}-\alpha_n\right)\sin(\delta\Phi_n/2)\bigg], \notag\\
\Delta_ne^{i\phi_n} &= iw_n \sin\lambda_n \sin\rho_n e^{i\xi_n} \times \notag\\
\bigg[&\cos\left(\alpha_{n+1}+\alpha_n\right)\cos\left(\delta\Phi_n/2\right)\\
+i&\,\sin\left(\alpha_{n+1}-\alpha_n\right)\sin\left(\delta\Phi_n/2\right)\bigg], \notag
\end{align}\label{eq:eff_params}
\end{subequations}
where
\begin{gather}
\mu_n=V_z\sin (2 \alpha_n) \text{, } \Delta_{\textrm{ind},n}=V_z\cos (2 \alpha_n),\\
\bm{\lambda}_n = \lambda_n \left(\sin\rho_n \cos\xi_n,\: \sin\rho_n \sin\xi_n,\: \cos\rho_n\right)^T,
\end{gather}
and $\delta\Phi_n = \Phi_n-\Phi_{n+1}$.

It is possible to extract most of the parameters of the dot Hamiltonian from 
level spectroscopy, and then tune the effective Kitaev chain Hamiltonian to
the sweet spot. The tuning, however, becomes much simpler if two out of three of the dot 
linear dimensions are much smaller than the spin-orbit coupling length. Then
the
direction of spin-orbit coupling does not depend on the dot number, and as long
as the magnetic field is perpendicular to the spin-orbit field, the phase of
the prefactors in Eqs.~\eqref{eq:eff_params} becomes position-independent.
Additionally, if the dot size is not significantly larger than the spin-orbit
length, the signs of these prefactors are constant. This ensures that if
$\delta\Phi_n = 0$, the phase matching condition of the Kitaev chain is
fulfilled. Since $\delta\Phi_n = 0$ leads to both $t_n$ and $\Delta_n$ having a
minimum or maximum as a function of $\delta\Phi_n$, this point is
straightforward to find. The only remaining condition, $t_n = \Delta_n$ at
$\delta\Phi=0$, requires that $\alpha_n 
+ \alpha_{n+1} = \lambda_n$.

The above calculation leads to the following tuning algorithm:
\begin{enumerate}
\item Open all the QPCs, except for two contacting a single dot. By measuring 
conductance while tuning the gate voltage of a nearby gate, ensure that there
is a resonant level at zero bias. After repeating for each dot the condition
$\varepsilon_n = 0$ is fulfilled.
\item Open all the QPCs except the ones near a pair of neighboring dots. 
Keeping the gate voltages tuned such that $\varepsilon_n = 0$, vary the phase 
difference between the neighboring superconductors until the lowest resonant 
level is at its minimum as a function of phase difference, and the
next excited level at a maximum. This ensures that the phase tuning condition
$\phi_{n+1} - \phi_n - \theta_{n+1} - \theta_n = 0$ is fulfilled. Repeat for
every pair of neighboring dots.
\item Start from one end of the chain, and isolate pairs of dots like in the
previous step. In the pair of $n$-th and $n+1$-st dots tune simultaneously the
coupling of the $n+1$-st dot to the superconductor and the chemical potential
in this dot, such that $\varepsilon_{n+1}$ stays equal to 0. Find the values of
these parameters such that a level at zero appears in two dots when they are
coupled. After that proceed to the following pair.
\end{enumerate}

Having performed the above procedures, the coupling between all of the dots in
the chain is resumed, at which point we expect the system to be in a robust
topological phase, with two Majorana fermions located on the first and last
dots. In practice one can also resume the coupling gradually by, for instance,
isolating triplets of adjacent dots, making sure they contain a zero-energy
state, and making fine-tuning corrections if necessary, and so on.

\section{Testing the tuning procedure by numerical simulations}
\label{perform_tune}

We now test the tuning procedure by applying it to a numerical simulation
of a chain of three QDs in a 2DEG. The two-dimensional BdG Hamiltonian 
describing the entire system of the QD chain reads: 
\begin{equation}
\begin{split}
\mathcal{H}_\text{QDC}&=\left( \frac{p^2}{2m} + V(x,y)
\right)\tau_z+\frac{\alpha}{\hbar}
(\sigma_xp_y - \tau_z\sigma_yp_x)\\
&+\Delta_\text{ind}
\left(\cos(\Phi)\tau_y+\sin(\Phi)\tau_x\right)\sigma_y+V_z\tau_z\sigma_z.\label
{ eq:Hqdc}
\end{split}
\end{equation}
Here, $\sigma_i$ and $\tau_i$ are
Pauli matrices acting on the spin and 
particle-hole degrees of freedom respectively. The term $V(x, y)$ describes
both potential fluctuations due to disorder, and the confinement potential
introduced by the gates. The second term represents Rashba spin-orbit coupling,
$\Delta_\text{ind}(x,y)\cdot\exp\left(\Phi(x,y)\right)$ is the s-wave
superconductivity induced by the coupled superconductors, and $V_z$ is the
Zeeman splitting due to the magnetic field. Full description of the
tight-binding equations used in the simulation is presented in Appendix
\ref{tuning}.

The chemical potential of the dot levels $\mu_n$ is tuned by changing the
potential $V(x,y)$. For simplicity we used a constant potential $V_n$ added to 
the disorder potential, such that $V(x,y) = V_n + V_0(x,y)$ in each dot.
Varying the magnitude of $\Delta_{\text{ind},n}$ is done by changing
conductance $G_\perp$ of the quantum point contacts, which control the coupling
between the dots and the superconductors. Finally, varying the superconducting
phase $\Phi(x, y)$
directly controls the parameter $\Phi_n$ of the dot to which the superconductor is coupled, although they need not be the same.

The tuning algorithm required monitoring the energy levels of different parts of the system. This can be achieved by measuring the resonant Andreev conductance from one of the leads. The Andreev conductance is given by\cite{She80, Blo82}
\begin{equation}
    G/G_0 = N - \tr(r_{ee}r^\dag_{ee}) + \tr(r_{he}r^\dag_{he})\label{eq:cond},
\end{equation}
where $G_0=e^2/h$, $N$ is the number of modes in a given lead, and $r_{ee}$ and $r_{he}$ are
normal and Andreev reflection matrices.
Accessing parts of the chain (such as a single dot or a pair of dots) can be done by opening all inter-dot QPCs, and closing all the ones between dots and superconductors, except for part of the system that is of interest.

We begin by finding such widths of QPCs that $G_\parallel \approx 0.02$ and $G_\perp \approx 4 G_0$.
This ensures that conductance between adjacent dots, is in tunneling regime and that the dots are strongly 
coupled to the superconductors such that the effect of Coulomb blockade is reduced.\cite{Gra92} The detailed properties of QPCs are descrbed in App.~\ref{tuning} and their conductance is shown in Fig.~\ref{fig:qpc}.

\begin{figure}[tb]
 \begin{center}
   \includegraphics[width=0.8\linewidth]{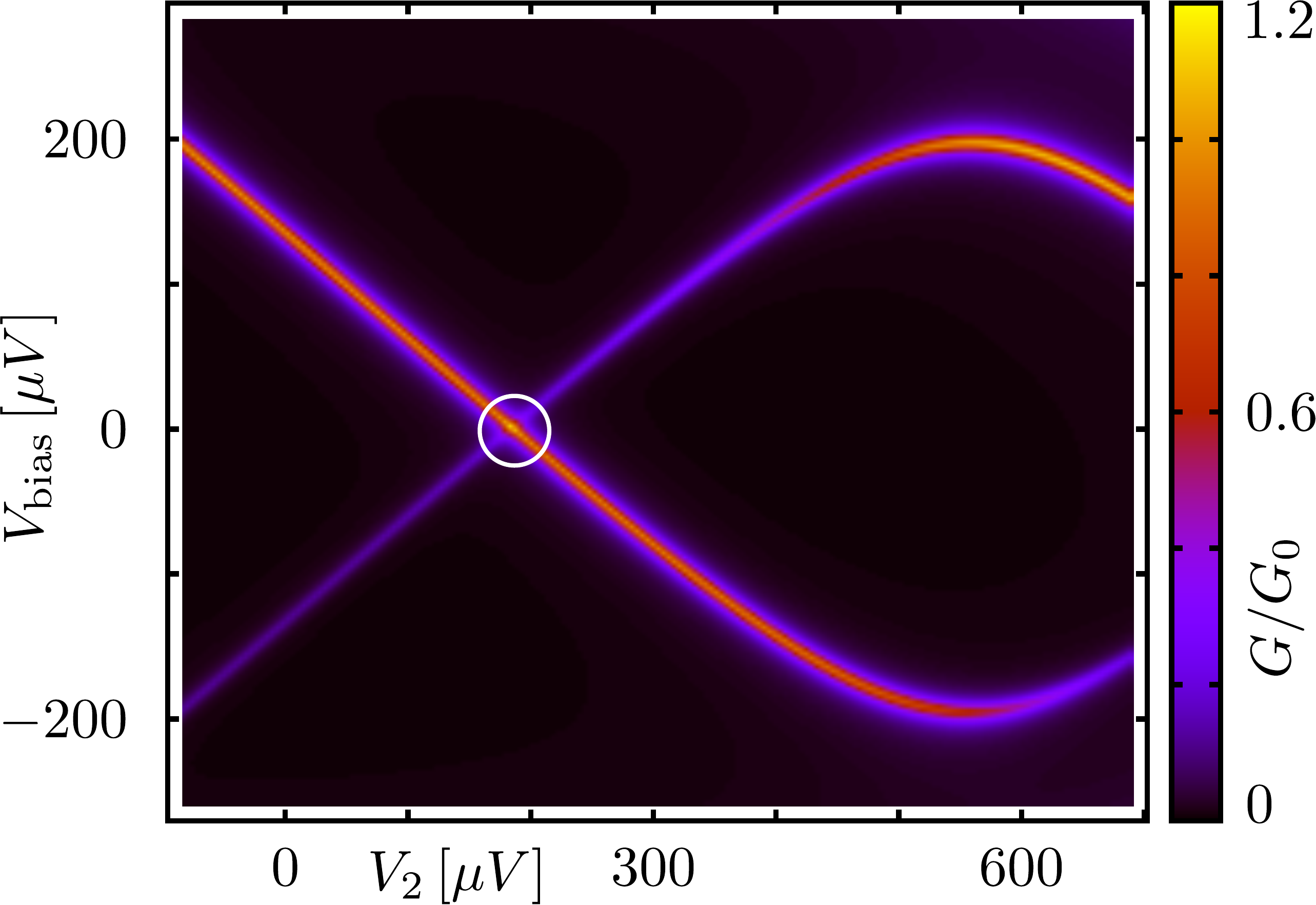}
  \caption{Andreev conductance measured from the left lead as a function of
bias voltage and QD potential (measured relative to quarter filling) for
the second dot. Changing the chemical potential allows to tune quasi-bound
states to zero energy (white circle).
\label{fig:singledot}}
 \end{center}
\end{figure}

\emph{First step: tuning chemical potential.}
We sequentially isolate each dot, and change the dot potential $V_n$. The Andreev
conductance as a function of $V_n$ and bias voltage for the second dot is
shown in Fig.~\ref{fig:singledot}. We tune $V_n$ to the value where a
conductance resonance exists at zero bias. This is repeated for each of the
dots and ensures that $\mu_n=0$.

\begin{figure}[tb]
 \begin{center}
   \includegraphics[width=0.8\linewidth]{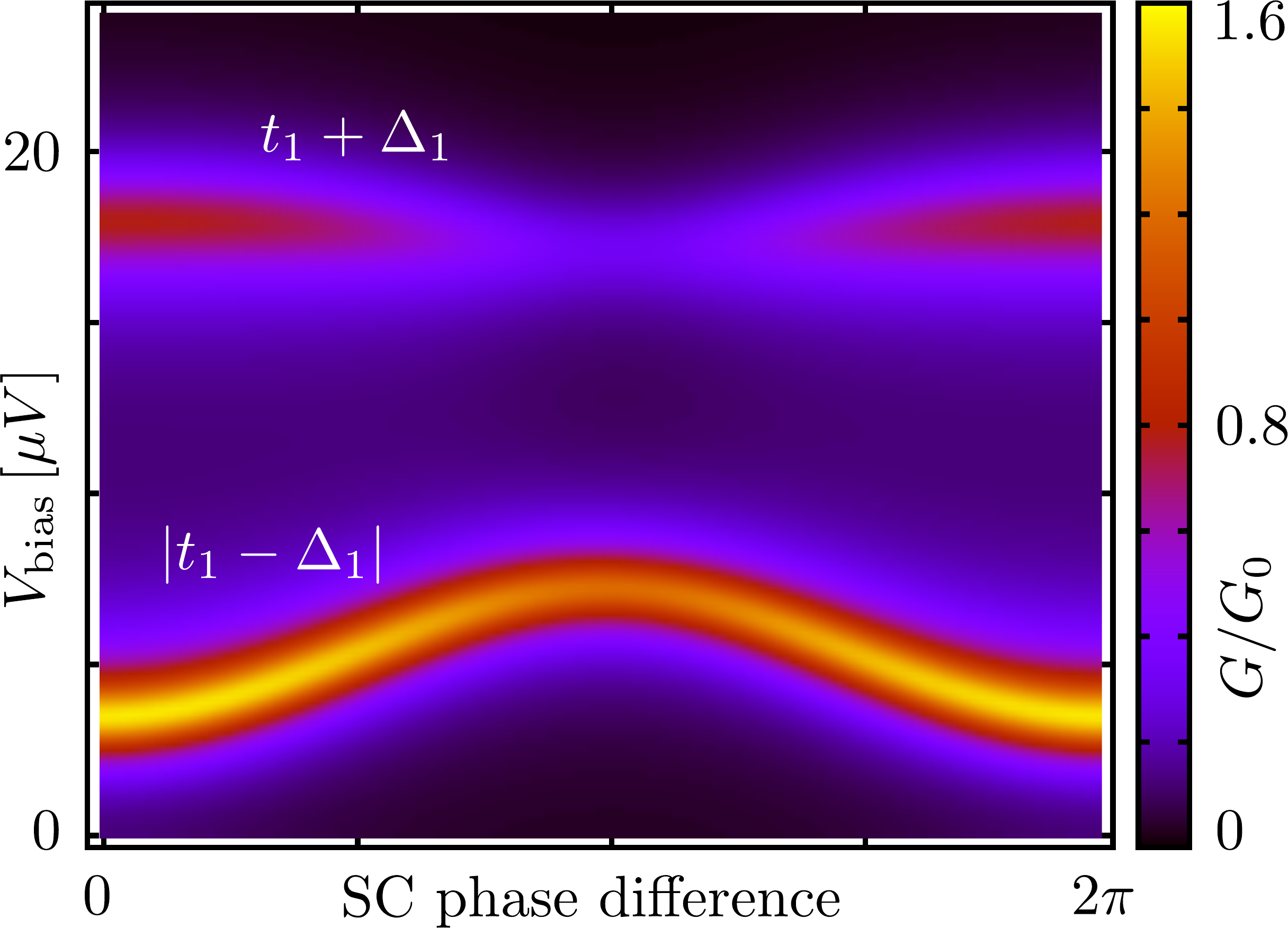}
  \caption{Conductance as a function of bias voltage and superconducting
phase difference for a two-dot system. The two lowest energy levels are given
by Eq.~\eqref{eq:twosites} of a two site Kitaev chain, as indicated. At the
point where their separation is maximal (SC phase difference 0 in the plot), 
the phase difference $\delta \Phi_n$ of the induced superconducting gaps vanishes.
\label{fig:2dot_levels}}
 \end{center}
\end{figure}

\emph{Second step: tuning the superconducting phases.}
We now set the phases of the induced pairing potentials $\Phi_n$ to constant.
As explained in the previous section, this occurs when $\Delta_n$ and $t_n$
experience their maximal and minimal values. According to Eq.~\eqref{eq:twosites}
this happens when the separation between the energy levels of the pair of dots
subsection is maximal. Fig.~\ref{fig:2dot_levels} shows the evolution of these
levels as a function of the phase difference between the two superconductors.
The condition $\delta\Phi_1=0$ is then satisfied at the point where
their separation is maximal.

\begin{figure}[tb]
 \begin{center}
   \includegraphics[width=0.9\linewidth]{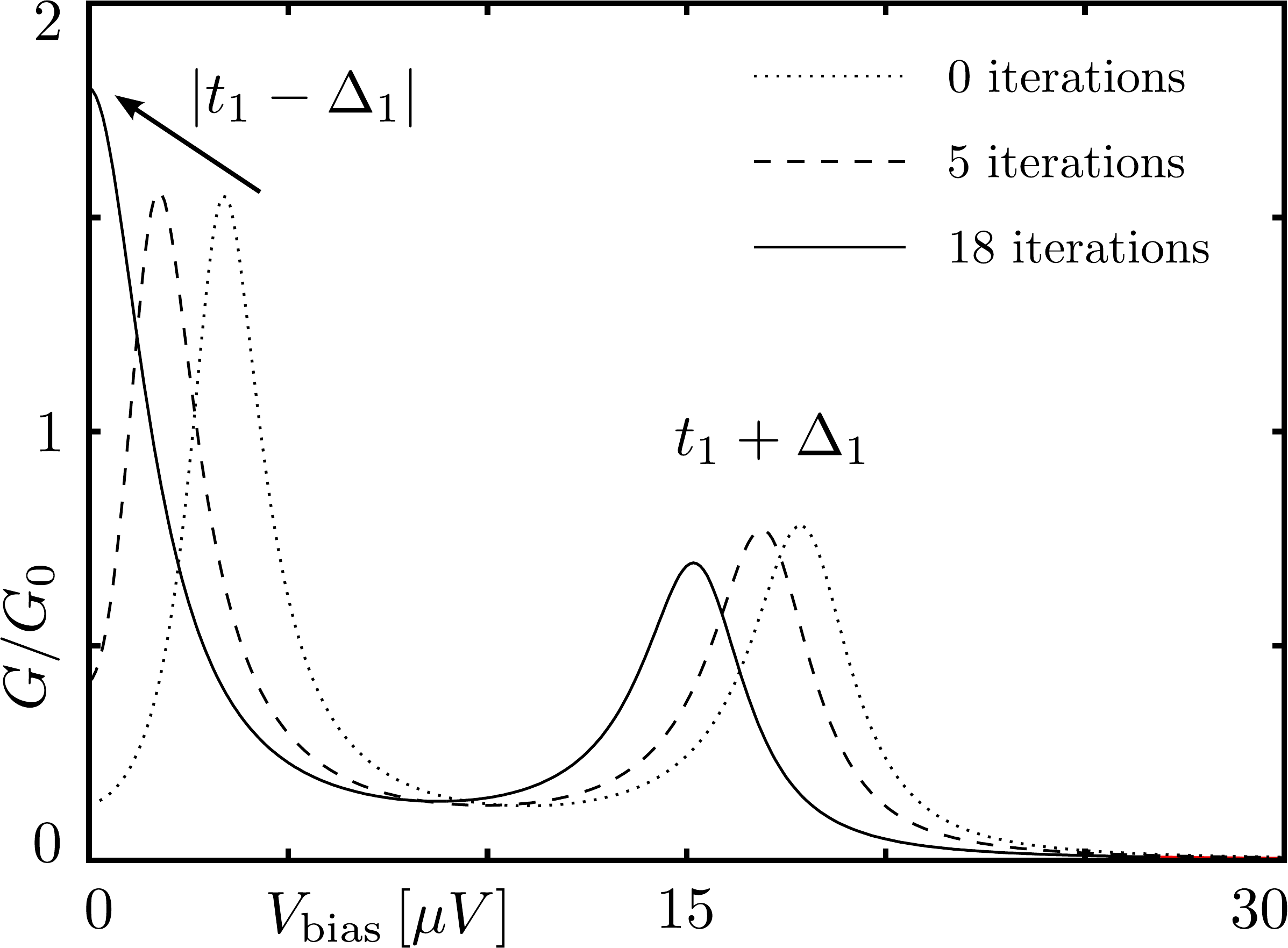}
  \caption{Conductance as a function of bias voltage during simultaneous tuning
of $G_\perp$ and $V_n$ for the first pair of dots. The three
different plots represent the situation before (dotted line), at an
intermediate stage (dashed line), and after (solid line) the tuning. The
arrow indicates the evolution of the first peak upon tuning, and the number of
successive changes of $G_\perp$ and $V_n$ are shown for each curve.
By bringing the first peak to zero, the third tuning step is achieved.
\label{fig:twodots_tuning}}
 \end{center}
\end{figure}

\emph{Third step: tuning the couplings.}
Finally we tune $t_n=\Delta_n$. This is achieved by varying $G_\perp$, while tracking the Andreev conductance peak corresponding to
the $t_n - \Delta_n$ eigenvalue of the Kitaev
chain we are emulating. After every change of $G_\perp$ we readjust
$V_n$ in order to make sure that the condition $\varepsilon_n=0$ (or equivalently $V_z^2 = \mu_n^2 + \Delta_n^2$) is
maintained. This is necessary because not just $\Delta_n$, but also $\mu_n$
depend on $G_\perp$. Therefore, successive changes of $G_\perp$ and $V_n$ are
performed until the smallest bias peak is located at zero bias.
The tuning steps of the first two dots are shown in
Fig.~\ref{fig:twodots_tuning}. We repeat steps 2 and 3 for each pair of dots
in the system.

\begin{figure}[tb]
 \begin{center}
   \includegraphics[width=0.9\linewidth]{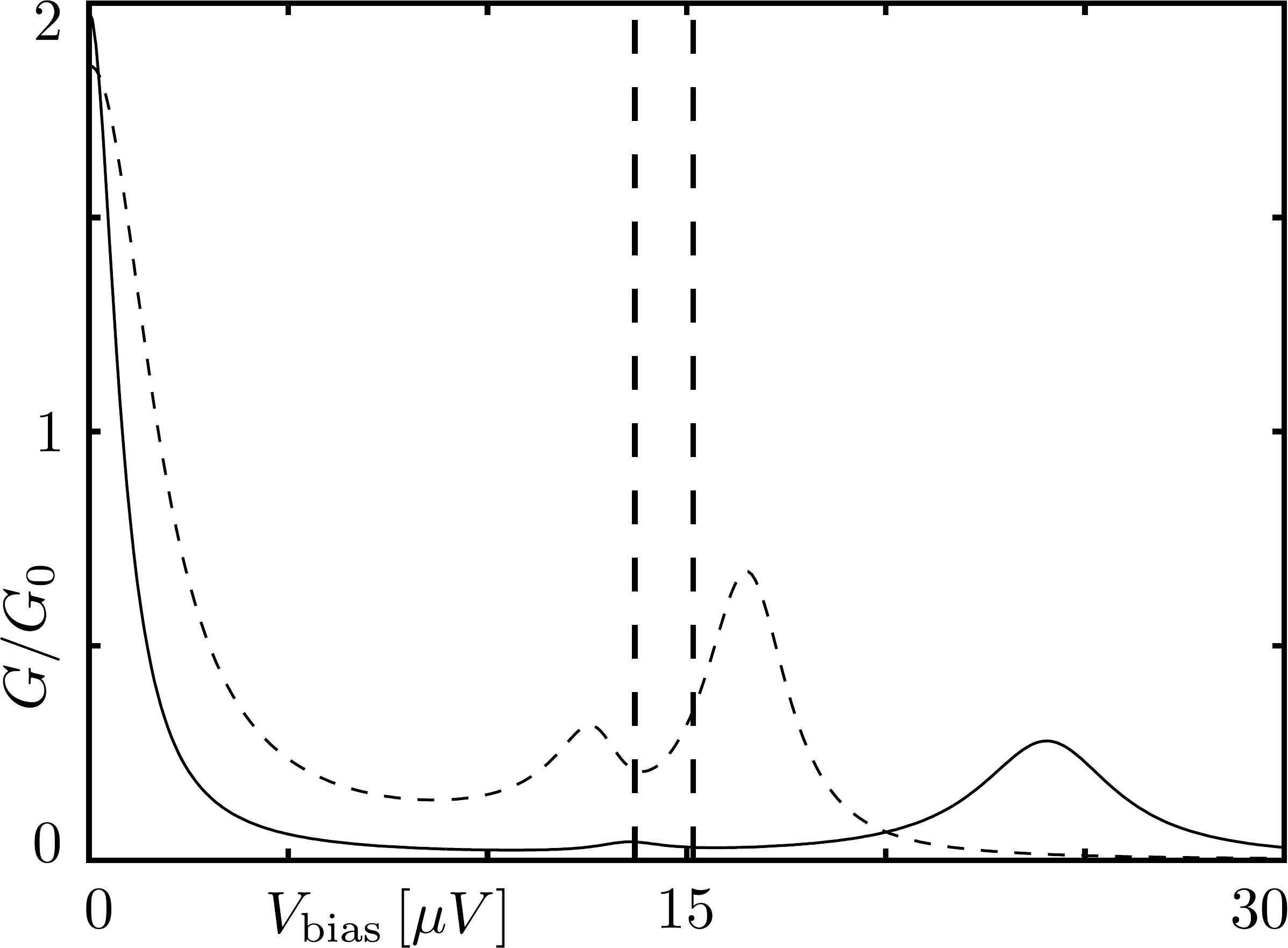}
\caption{Conductance as a function of bias voltage for a system composed of
three tuned quantum dots (dashed line). The zero bias peak signals the presence
of Majorana bound states at the ends of the chain. The first and second excited
states are consistent with those expected for a three-site Kitaev chain, namely
$E_1=2t_1$ and $E_2=2t_2$ (vertical dashed lines), given the measured values of
$t_1=\Delta_1$ and $t_2=\Delta_2$, obtained after finalizing the two dot tuning
process. As described in the main text, after increasing the transparency of
the lead QPC leads to a zero bias peak having a height $G = 1.98 G_0$ (solid
line).
\label{fig:threedots}}
 \end{center}
\end{figure}

Finally, having full all three conditions required for a robust
topologically non-trivial phase, we probe the presence of localized Majorana bound state in
the full three-dot system by measuring Andreev conductance
(see Fig.~\ref{fig:threedots}). In this specific case, the height of the zero
bias peak is approximately $1.85\,G_0$, signaling that the end states are well
but not completely decoupled. Increasing the transparency of the QPC
connecting the first dot to the lead brings this value to $G = 1.98\,G_0$.

\section{Conclusion}
\label{conc}

In conclusion, we have demonstrated how to tune a linear array of quantum dots
coupled to superconductors in presence of Zeeman field and spin-orbit coupling
to resemble the Kitaev chain that hosts Majorana bound states at its ends.
Furthermore, we have presented a detailed procedure by which the system is
brought to the so-called ``sweet spot'' in parameter space, where the Majorana
bound states are the most
localized. This procedure involves varying the gates potentials and
superconducting phases, as well as monitoring of the excitation spectrum of the
system by means of resonant Andreev conductance.

We have tested our procedure using numerical simulations of a system of three
QDs, defined in a 2DEG, and found that it works in systems with experimentally
reachable parameters. It can be also applied to systems where quantum dots are
defined by other means, for example formed in a one-dimensional InAs or InSb
wire.

\acknowledgments
The numerical calculations were performed using the {\sc kwant} package developed by A. R. Akhmerov, C. W. Groth, X. Waintal, and M. Wimmer. We would like to acknowledge discussions with J. Alicea, L. P. Kouwenhoven, C.
M. Marcus, F. von Oppen, and J. D. Sau. We are grateful for partial support by
SPP 1285 of the Deutsche Forschungsgmeinschaft (YO), for grants of ISF and
TAMU (YO), to the Dutch Science Foundation NWO/FOM and an ERC Advanced
Investigator Grant (ICF and AA). AA was partially supported by a Lawrence Golub
Fellowship.

\appendix

\section{System parameters in numerical simulations}
\label{tuning}

In this section, we describe the parameters used throughout the numerical
simulations. The quantum dots and quantum point contacts are modeled using a
tight-binding model defined on a square lattice, with leads and
superconductors taken as semi-infinite.

The characteristic length and energy scales of this system are the
spin-orbit length $l_\text{SO} = \hbar^2/m\alpha$, and the
spin-orbit energy $E_\text{SO}=m\alpha^2/\hbar^2$. We simulate an InAs system
in which the effective electron mass is $m = 0.015 m_e$, where $m_e$ is the
bare electron mass, taking values of $E_\text{SO} = 1\,\text{K}= 86\,\mu
eV$ and $l_\text{SO} = 250\,\text{nm}$.

\begin{figure}[tb]
 \begin{center}
   \includegraphics[width=0.9\linewidth]{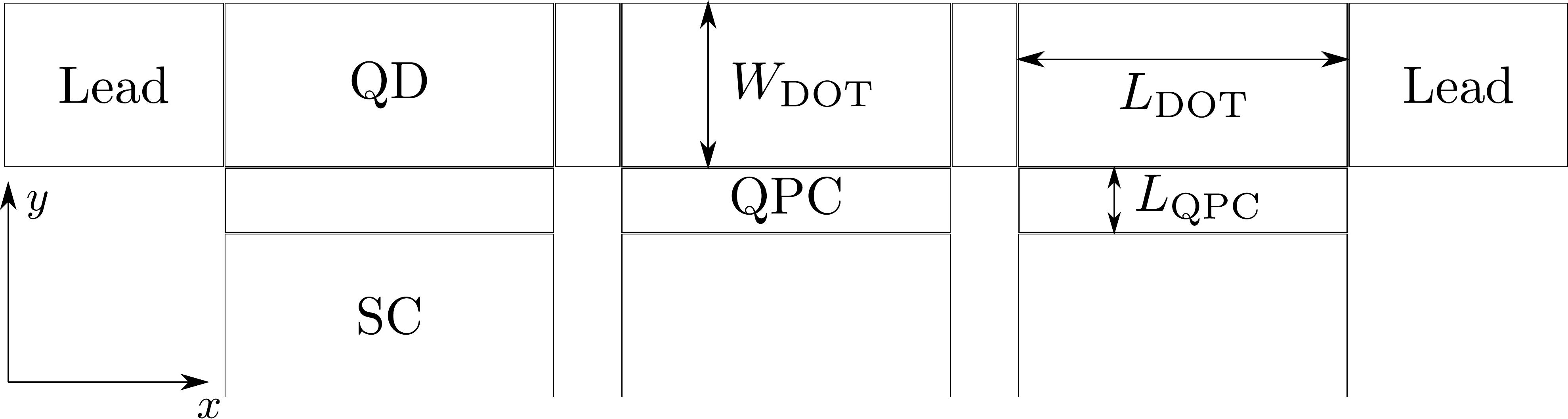}
\caption{Geometry of the quantum dot chain. The quantum dots have a width
$W_{\rm DOT}$ and length equal to $L_{\rm DOT}$. Quantum point contacts have a
longitudinal size $L_{\rm QPC}$ and a transverse dimension equal to
either $L_{\rm DOT}$ or $W_{\rm DOT}$. Leads are semi-infinite in the $x$
direction, and superconductors are modeled as semi-infinite systems in the $y$
direction.
\label{fig:dots_scheme}}
 \end{center}
\end{figure}

We consider a setup composed of three quantum dots, like the one shown in
Fig.~\ref{fig:dots_scheme}. Each of the three dots has a length of $L_{\rm DOT}
= 208\,\text{nm}$ and a width $W_{\rm DOT} = 104\,\text{nm}$. Quantum
point contacts have a longitudinal dimension of $L_{\rm QPC} = 42\,\text{nm}$,
which is the same as the Fermi wavelength at quarter filling.

The value of the hopping integral becomes $t=\hbar^2 / (2ma^2) = 55.8\,
\text{meV}$, with $a=7\,\text{nm}$. Disorder is introduced in the form of
random uncorrelated onsite potential fluctuations, leading to a mean free path
$l_\text{mfp} = 218.8\,\text{nm}$. The system is placed in a perpendicular
magnetic field characterized by a Zeeman splitting $V_z = 336\,\mu\text{eV}$,
which, given a $g$-factor of $35 K/T$, corresponds to a magnetic field $B_z =
111\,\text{mT}$. Each dot is additionally connected to a superconductor
characterized by a pairing potential $|\Delta_\text{SC}| = 0.86\,\text{meV}$.

\begin{figure}[htb]
 \begin{center}
   \includegraphics[width=0.9\linewidth]{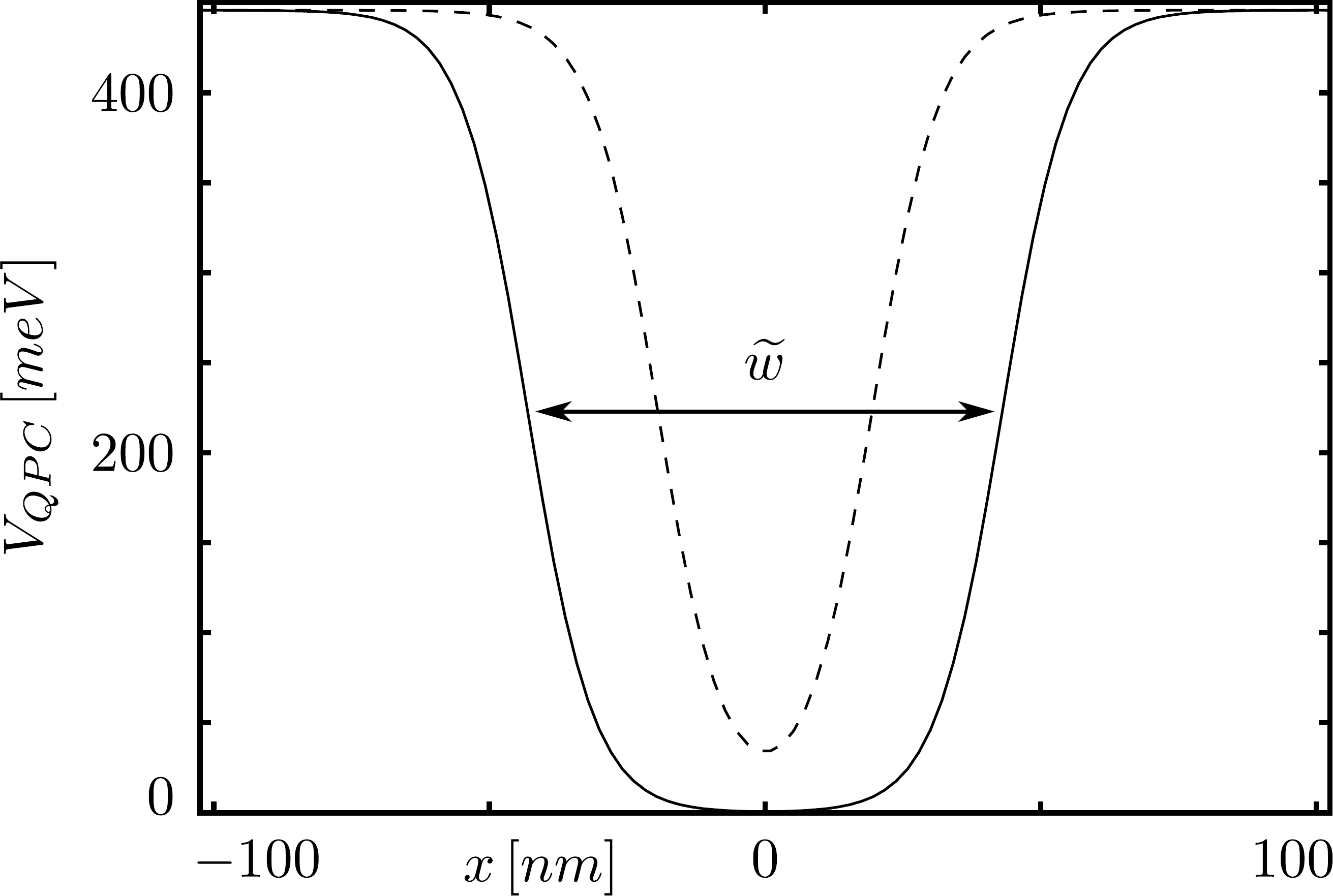}
\caption{Potential profile $V_{\rm QPC}(x)$ across the transverse direction of
a quantum point contact. For the maximum value of this potential, no states are
available for quasiparticles in the 2DEG. The two curves show potential
profiles for two different QPC transparencies, corresponding to
$\widetilde{s}=17$ and $\widetilde{w} = 87.4$, $39.5$ nm for the
solid and dashed curves respectively.
\label{fig:qpc_potential}}
 \end{center}
\end{figure}

The potential profile across a quantum point contact is given by
\begin{eqnarray}
    V_{\rm QPC}(x) &=& \frac{\widetilde{h}}{2}  \left(2-\tanh
\left(\frac{\widetilde{s}}{\widetilde{L}} \left( x+\frac{\widetilde{w}}{2}
\right) \right) \right. \nonumber\\
 & & \left. +\tanh \left({ \frac{\widetilde{s}}{\widetilde{L}} \left(
x-\frac{\widetilde{w}}{2} \right)} \right)
\right), \label{eq:vqpc}
\end{eqnarray}
where $x \in [-\widetilde{L}/2, \widetilde{L}/2]$ is the transverse
coordinate across the quantum point contact, $\widetilde{h}$ is the maximum
height of $V_{\rm QPC}$, $\widetilde{s}$ fixes the slope at which the potential
changes, and $\widetilde{w}$ is used to tune the QPC transparency. Two examples
of potential profiles are shown in Fig.~\ref{fig:qpc_potential}.

\begin{figure}[htb]
 \begin{center}
   \includegraphics[width=0.8\linewidth]{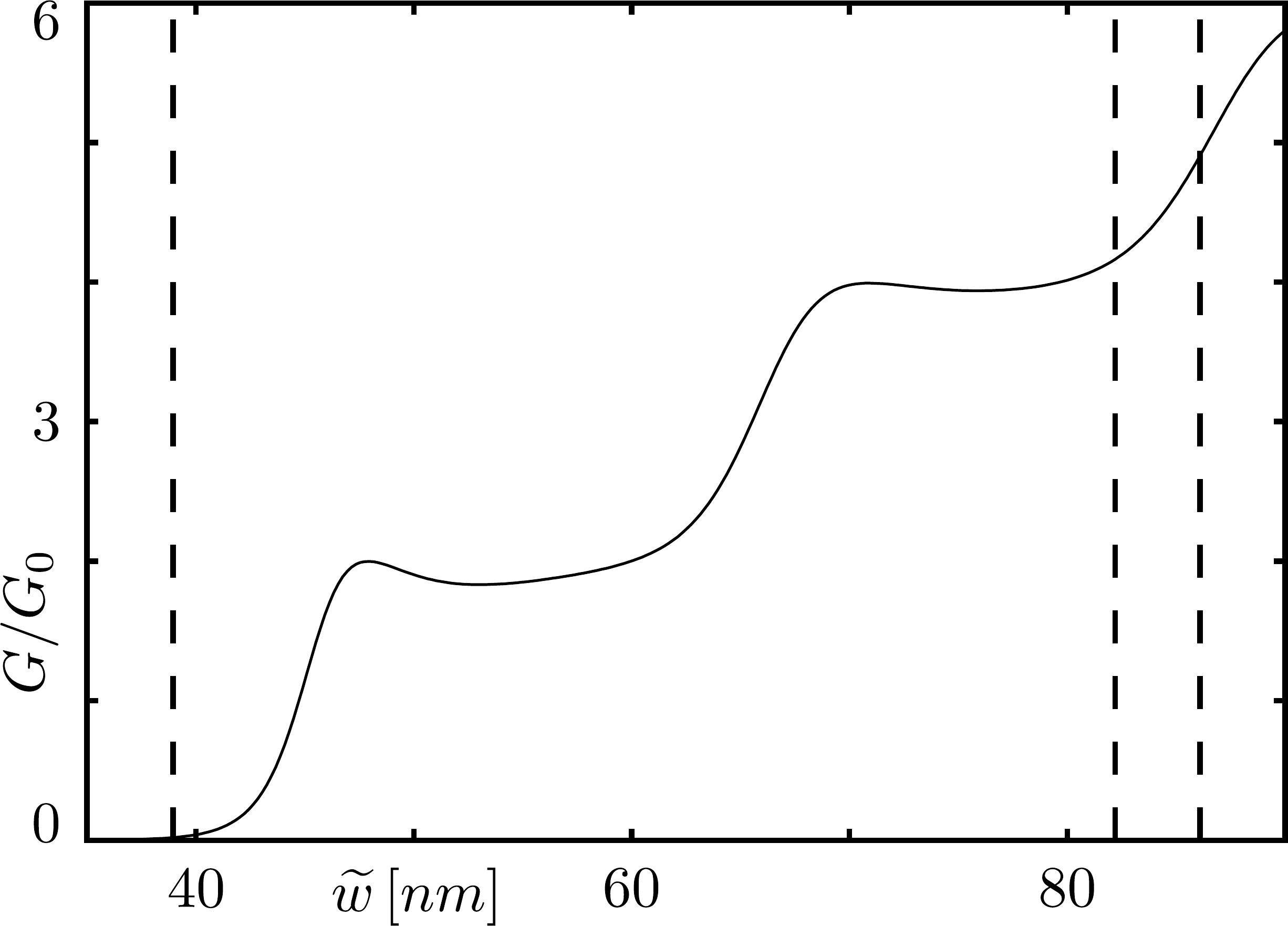}
  \caption{Conductance of a quantum point contact as a function of
$\widetilde{w}$ of Eq.~\eqref{eq:vqpc}, for a single QPC. The
vertical lines indicate the values at which QPCs are set after tuning.
The inter-dot QPCs are all set to the tunneling regime while the
ones connecting the dots to the superconductors are set to higher
transparencies.\label{fig:qpc}}
 \end{center}
\end{figure}

\end{document}